\pdfoutput=1
\documentclass[final,3p,times]{elsarticle}
\usepackage{amssymb}
\usepackage{palatino}
\usepackage{graphicx}
\usepackage{epsfig}
\usepackage{booktabs}
\usepackage{array}
\usepackage{paralist}
\usepackage{verbatim}
\usepackage{subfig}
\usepackage{dsfont}
\usepackage{bibentry}
\usepackage{hyperref}
\usepackage{color}
\usepackage{tikz}
\usepackage{multirow}
\usepackage{amsthm}
\usepackage{amsmath, amsfonts,amssymb, amsthm}
\usepackage{algorithm}
\usepackage{algorithmicx}
\usepackage{algpseudocode}
\usepackage{graphicx,epstopdf}
\usepackage{xspace}
\usepackage{float}

\newtheorem{theorem}{Theorem}[section]
\journal{arXiv}

\begin{document}

\begin{frontmatter}


\title{A Data-Driven Model-Free Physics-Informed Deep Operator Network for Solving Nonlinear Dynamic Systems}

\author[1]{Jieming Sun\corref{cor1}}
\ead{js18hc@fsu.edu}

\author[1]{Lichun Li}
\ead{lichunli@eng.famu.fsu.edu}

\cortext[cor1]{Corresponding author}

\affiliation[1]{organization={Department of Industrial and Manufacturing Engineering, FAMU-FSU College of Engineering},
                addressline={Florida State University},
                city={Tallahassee},
                postcode={32306},
                state={FL},
                country={USA}}

\fntext[label1]{Both authors contributed equally to this work.}
%\fntext[label3]{}


\begin{abstract}
The existing physical-informed Deep Operator Networks are mostly based on either the well-known mathematical formula of the system or huge amounts of data for different scenarios.  However, in some cases, it is difficult to get the exact mathematical formula and vast amounts of data in some dynamic systems, we can only get a few experimental data or limited mathematical information. To address the cases, we propose a data-driven model-free physical-informed Deep Operator Network (DeepOnet) framework to learn the nonlinear dynamic systems from few available data. We first explore the short-term dependence of the available data and use a surrogate machine learning model to extract the short-term dependence.  Then, the surrogate machine learning model is incorporated into the DeepOnet as the physical information part. Then, the constructed DeepOnet is trained to simulate the system's dynamic response for given control inputs and initial conditions. Numerical experiments on different systems confirm that our DeepOnet framework learns to approximate the dynamic response of some nonlinear dynamic systems effectively.
\end{abstract}


\end{frontmatter}


\section{Introduction}
\label{sec:intro}


In the vast realm of dynamic systems analysis, estimating and anticipating states in dynamic systems stands as a cornerstone in numerous scientific and engineering domains. This pivotal endeavor involves deducing an evolving signal based on its historical observations—even when these observations may deviate from the signal being estimated—alongside any understood physical principles governing its progression \cite{close2001modeling}. 

The increasing complexity of these systems has ignited substantial interest in the scientific community to conceive tools that can expedite the numerical simulation of dynamic systems. In a groundbreaking development, the Deep Operator Network (DeepONet)\cite{lu2021learning} was proposed as a framework to learn nonlinear operators, essentially facilitating mappings from one functional space to another. The foundational principle behind DeepONet is based on the universal approximation theorem of operators, which postulates that nonlinear operators can be approximated with the assistance of scattered data streams. DeepONet has been tweaked to include a physics-based understanding by using a loss function inspired by the Physics-informed neural network \cite{karniadakis2021physics}. This gave rise to the Physical-informed DeepOnet \cite{wang2021learning}, expanding its usefulness in various applications.

Furthermore, a myriad of researchers have employed DeepONet for analyzing or learning dynamic systems, as demonstrated in works like \cite{lin2023learning}, \cite{wang2021modeling}, and \cite{garg2022assessment}. Yet, a common thread among these studies is the reliance on existing Deep Operator Networks or their derivatives, which predominantly lean on either established mathematical formulas of the system or huge amounts of data. Regrettably, in certain instances, procuring a precise mathematical formula or extensive datasets for dynamic systems becomes a daunting task. We might find ourselves restricted to a handful of experimental data points or scant mathematical insights.

In this paper, we address the challenge of offline state estimation in dynamic systems operating under constraints of limited information. This constraint pertains both to the availability of data and the comprehensive understanding of the system's physical rules or equations. In certain scenarios, acquiring an exact mathematical representation or formula of the system's dynamics proves to be a challenge \cite{bar2002general}\cite{bishop2011modern}. Our grasp of the system model might be restricted, knowing, for instance, that the system can be defined by an Ordinary Differential Equation (ODE) of a particular order, but lacking specificity in the formula itself. Data constraints further compound this challenge. Experimental endeavors, vital for understanding and mapping these systems, often come with significant costs \cite{thon2023financial}, thus limiting the number of feasible experiments.

To tackle the challenges presented, we introduce a data-driven, model-free, physics-informed Deep Operator Network (DeepOnet) approach. This approach focuses on deciphering nonlinear dynamic systems even when only limited data and limited system information is accessible. Section \ref{sec:preli} sheds light on the foundational concepts required for our discussion, encapsulating introductions to PINN, DeepOnet, and the principle of short-term dependency. Our methodology takes center stage in section \ref{sec:method}. Here, we first delve into the short-term dependence present in the available data. Subsequently, we employ a surrogate machine learning model to capture this dependence. This model is then seamlessly integrated into DeepOnet, serving as its physics-informed component. Following this, we train the developed DeepOnet, directing it to emulate the system's dynamic reactions based on specified control inputs and starting conditions. Our exploration culminates in section \ref{sec:result}, where, through numerical tests on diverse systems, we demonstrate the proficiency of our DeepOnet model in closely approximating the dynamic responses of certain nonlinear dynamic systems.

\section{Preliminary}
\label{sec:preli}

In this section, we provide an overview of Physics-informed Neural Networks (PINN), Deep Operator Neural Networks (DeepOnet), and short-term dependency, all of which are utilized in our study.

\subsection{Physical-informed Neural Networks}
\label{subsec:PINN}

In recent times, Physics-informed Neural Networks (PINN) have emerged as a noteworthy advancement in numerical computing, captivating the interest of researchers across the domains of science and engineering \cite{karniadakis2021physics}. This method, initially introduced by Karniadakis and his team \cite{raissi2019physics, raissi2017physics,raissi2018deep}, leverages machine learning to train neural networks, aiming to approximate solutions to partial differential equations (PDEs) and similar physics-oriented challenges.

What differentiates PINNs is their ability to integrate physical conservation principles and established physical knowledge into their architecture. This not only guarantees the precise depiction of underlying physics but also lessens the reliance on traditional supervised learning. With PINNs, it becomes feasible to predict system state variables for any temporal instance. Moreover, these networks exhibit resilience against data imperfections, like noise or omissions, and consistently produce predictions that resonate with established physical truths. This is attributed to their capability to process available data in harmony with any physical laws defined by nonlinear PDEs. A key feature of PINNs is their fusion of data from both measurements and PDEs, achieved by assimilating the PDEs into the loss function of the neural network through automatic differentiation.

Consider a neural network function $\hat{y}(x_i,t_i,\theta)$ with specific activation functions and a weight matrix $\theta$. This matrix $\theta$ represents the degrees of freedom determined by the network's width and depth. The goal is to identify $\theta$ that minimizes the loss function.

The total loss of the PINN combines the data loss $L_{D}$ and the physical loss $L_{F}$, defined as:
\begin{equation}
L = w_{D}L_{D} + w_{F}L_{F}
\end{equation}

Here, the data loss $L_{D}$ measures the discrepancy between the neural network's predicted value $y(x_i,t_i,\theta)$ and the actual measurement $\bar{y}$ from the initial and boundary conditions. Using the mean squared error (MSE) loss, it is expressed as:
\begin{equation}
L_{D}= MSE_D = \frac{1}{N_{D}}\sum_{i=1}^{N_{D}}|(x_i,t_i,\theta)-\bar{y}_i|^2
\end{equation}

The physical loss, denoted as \( L_{F} \), serves as an unsupervised metric, grounded in a distinct physical rule depicted by \( F() \). It is quantified using the MSE, which we label as \( MSE_F \). The expression for \( L_{F} \) is:

\begin{equation}
L_{F} = MSE_F = \frac{1}{N_{F}}\sum_{j=1}^{N_{F}}|F(\hat{y}(x_j,t_j,\theta))|^2 \Bigg|_{(x_j,t_j)}
\end{equation}

Where $(x_i,t_i)$ and $(x_j,t_j)$ are points sampled from initial/boundary locations and across the entire domain, respectively. The weights $w_{D}$ and $w_{F}$ balance the data and physical losses. Their values can be preset or adjusted dynamically, playing a pivotal role in enhancing the effectiveness of PINNs. 

This loss-function modified minimization approach fits naturally into the traditional deep learning framework \cite{kelleher2019deep}. The network is trained by minimizing the loss function using various optimization algorithms such as Adam\cite{kingma2014adam} or L-BFGS \cite{byrd1995limited}. During training, the optimizer adjusts $\theta$ the set of weights and biases in the neural network to minimize the loss function. The training process continues until the loss function reaches a pre-defined threshold, indicating that the network has learned an accurate approximation of the solution that satisfies the physical constraints.

The strength of PINN lies in its ability to handle sparse data by integrating physical principles. As highlighted in section \ref{sec:intro}, our aim is to address challenges posed by limited data, and in this context, PINN proves instrumental. However, we encounter another hurdle: the restricted system model information prevents us from accessing the exact formulation of the physical rules. Consequently, the conventional PINN approach might not be fully applicable. In this study, we substitute the physical rule with the concept of short-term dependency, as elaborated in section \ref{subsec:short}.

\subsection{Deep Operator Network }
\label{subsec:deepo}

DeepOnet is a type of neural network architecture proposed by Lulu and co-workers in the seminal paper \cite{lu2021learning} as a framework to learn nonlinear operators (i.e., mappings from a function space to another function space). They designed DeepOnet based on the universal approximation theorem of operators \cite{chen1995universal}.

The Universal Approximation Theorem is a fundamental result in the field of neural networks and machine learning. It is widely known that neural networks are universal approximators of continuous functions. A less known and perhaps more powerful result is that a neural network with a single hidden layer containing a sufficient number of neurons can approximate any continuous function on a compact subset of Euclidean space. In other words, given any function that maps inputs to outputs, a neural network architecture exists with a single hidden layer that can approximate the function to any desired level of accuracy. DeepOnet extends the Universal Approximation Theorem to deep neural networks (DNNs) with small generalization errors. DeepOnet also defines a new and relatively under-explored realm for DNN-based approaches that map infinite-dimensional functional spaces rather than finite-dimensional vector spaces (functional regression).

\begin{figure}[H]
 	\includegraphics[scale=0.6]{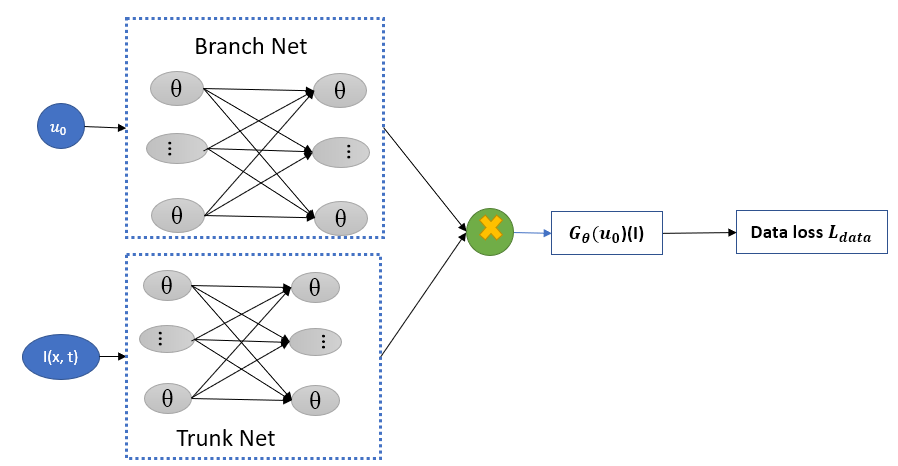}
 	\caption{The DeepOnet architecture}
 	\label{fig:DeepOnet}
 \end{figure}
 
DeepOnet architecture consists of two neural networks: one encodes the input function at fixed sensor points (branch net), while another encodes the information related to the spatial-temporal coordinates of the output function (trunk net). In general, the input $u$ to the branch net is flexible, i.e., it can take the shape of the physical domain, the initial or boundary conditions, constant or variable, source terms, etc.  The trunk net takes the spatial and temporal coordinates $y$ as the input.  

The goal of the DeepOnet is to learn the solution operator $G()$ that can be evaluated at spatial-temporal coordinates, $I$ (input to the trunk net). The output $G()$ of the DeepOnet for a specified input vector $u_0$, is a scalar-valued function of $I$ expressed as $G(u)(I)$ as
\begin{equation}
	G(u)(I) = \underbrace{b(u_0)}_{branch}\cdot\underbrace{t_k(I)}_{trunk}
\end{equation}

Similar to PINN, DeepOnet's trainable parameters, denoted by $\theta$, are derived by minimizing a loss function. DeepOnet primarily utilizes a data loss, which quantifies the discrepancy between the true value $\bar{y}$ and its prediction $G(u)(I)$ for the given input $(u_0, I)$. This loss, when calculated using the mean squared error (m.s.e.), can be represented as:
\begin{equation}
L_{D}= MSE_D = \frac{1}{N_{D}}\sum_{i=1}^{N_{D}}|G(u)(I_i)-\bar{y}_i|^2
\end{equation}.

DeepOnet's adaptability stems from its capability to utilize various branch and trunk networks like FNNs, CNNs, RNNs, and ResNets, allowing it to cater to diverse data types and challenges. It can discern both explicit operators, such as integrals, and implicit ones representing deterministic and stochastic differential equations. DeepOnets excel in rapid predictions of intricate dynamics for novel parameters while preserving accuracy due to training on vast input-output pair datasets. While they may not surpass the accuracy of numerical solvers rooted in mathematical principles, DeepOnets offer superior computational efficiency, making them faster, especially for complex problems. Furthermore, their data-driven learning enables them to discern and predict based on discerned data patterns.

DeepOnet offers versatility over PINN by accommodating a broader spectrum of operators, such as those emerging in PDEs, integral equations, and varied computational procedures. Yet, one challenge with DeepOnet is its substantial reliance on experimental data to comprehend the system's dynamics. In situations where data is sparse, the principles of PINN can be beneficial. PINN integrates known physical laws, providing an enhancement to DeepOnet's performance, especially when data is limited.

However, in the context of our study, the physical laws governing the system are elusive. This paper delineates our methodology to enhance the efficacy of DeepOnet under the constraints of limited data and scarce system information.

\subsection{Short-term Dependency }
\label{subsec:short}

Dynamic systems, often termed sequential models, exhibit pronounced long-term dependencies. This signifies that present outputs are influenced substantially by historical data. Recent findings demonstrate that both long-term memory models like recurrent neural networks (RNNs) and short-term memory models such as feed-forward neural networks (FFNNs) have analogous efficacies when approximating the dynamics of such systems. This similarity in performance propels many practitioners towards the more simplistic FFNNs, regardless of the vast memory benefits offered by RNNs. Both models maintain consistency in modeling sequential data \cite{kelchtermans2017hard,dauphin2017language,bai2018empirical, oord2018parallel}.

Our prior research \cite{sun2021short} aimed to decipher short-term dependencies in discrete-time dynamic systems. We meticulously examined systems defined by continuous timelines and values, trying to furnish analytical perspectives on their short-term dependencies. Our primary focus was a particular subset of nonlinear dynamic systems guided by ordinary differential equations (ODE), wherein only system variables are perceptible.

To illustrate, consider the class of nonlinear dynamic systems delineated below:

\begin{align}
\label{sys:n continuous}
\begin{cases}
\frac{d^n\boldsymbol{x}}{dt^n} = F(x,\frac{dx}{dt},\frac{d^{2}x}{dt^2},\dots,\frac{d^{n-1}x}{dt^{n-1}}, u)\\
y = x
\end{cases}
\end{align}

In this representation, $x$ denotes the system state, $u$ indicates the system input, and $y$ is the system output. The differential terms relate to the system state, while the function $F()$ embodies the system's governing principles. The system represented by equation (\ref{sys:n continuous}) encapsulates ordinary differential equations, which are fundamental in numerous mathematical, social, and scientific realms. These applications span from geometry and analytical mechanics to astronomy, climate predictions, epidemiological models, stock trend analysis, and more \cite{chicone2006ordinary} \cite{denis2020overview}.

When employing the Euler method \cite{burden2011numerical} to sample this continuous time system at intervals of $\delta$ seconds, the resulting discrete-time system adheres to the subsequent difference equation:
\begin{align}
	\label{sys:euler}
	\begin{cases}
		x(t+1) &= x(t) + \delta \frac{dx}{dt}(t)\\
		\frac{dx}{dt}(t+1) &= \frac{dx}{dt}(t) + \delta \frac{d^{2}x}{dt^{2}}(t)\\
		&\vdots\\
		\frac{d^{n-1}x}{dt^{n-1}}(t+1) &= \frac{d^{n-1}x}{dt^{n-1}}(t) + \delta \frac{d^{n}x}{dt^{n}}(t)\\
		\frac{d^{n}x}{dt^{n}}(t+1) &= \frac{d^{n}x}{dt^{n}}(t) + \delta F(...)\\
		y(t) &= x(t)
	\end{cases}
\end{align}

where $\delta $ is the sampling period, we assume that $\delta$ is small enough such that the difference between the outputs of the approximated sampled system (\ref{sys:euler}) and the true sampled system is negligible.

\begin{theorem}
	\label{theorem:n step dependency}
	The approximated sampled discrete time system (\ref{sys:euler}) is $n$-step dependent, i.e. there exists a function $\gamma$ mapping from $x(t-n),\ldots,x(t-1)$ and $u(t-n),\ldots,u(t-1)$ to $x(t)$.
\end{theorem}

Theorem \ref{theorem:n step dependency} posits that when the sampling period is adequately minute, ensuring the Euler method provides a close approximation, the contemporary output exclusively hinges on the preceding $n$-step inputs and outputs, with $n$ representing the highest order of differential value. Remarkably, for the class of nonlinear dynamic systems defined in equation (\ref{sys:n continuous}), even in the absence of explicit knowledge of the governing function $F()$, we can still elucidate its short-term dependencies.

\section{Methodology}
\label{sec:method}

As outlined in Section \ref{sec:intro}, this paper's primary objective is to estimate system states while grappling with limited system model details and sparse experimental data. To be more precise regarding the system model information, the system's dynamics can be represented by an Ordinary Differential Equation (ODE) of a known order. Leveraging this known ODE order, combined with the preliminary insights on short-term dependency discussed in Section \ref{subsec:short}, enables us to discern the system's short-term dependencies.

\subsection{Using Machine Learning Models to Learn Short-Term Dependency}
\label{sec:machine}

In our proposed methodology, we integrate the system's short-term dependency into DeepOnet. This short-term dependency, viewed as an intrinsic system rule, is then incorporated into the network's loss function.  

We first explore the historical data needed to anticipate the subsequent output. The purpose of this step is to build a mapping from history data (initial condition, history control inputs, and history states) to the current state. When the system meets the criteria of the class of nonlinear dynamic systems, as represented in equation \ref{sys:n continuous}, emphasizing its highest $n_{th}$ order differential, it is inferred from Theorem\ref{theorem:n step dependency} that the system has an $n$-step dependency. 
This implies that a function, termed as $\gamma$, exists that connects past system states $x(t-n),\ldots,x(t-1)$ and preceding control actions $u(t-n),\ldots,u(t-1)$ to the current state $x(t)$.

\begin{figure}[H]
	\includegraphics[scale=0.47]{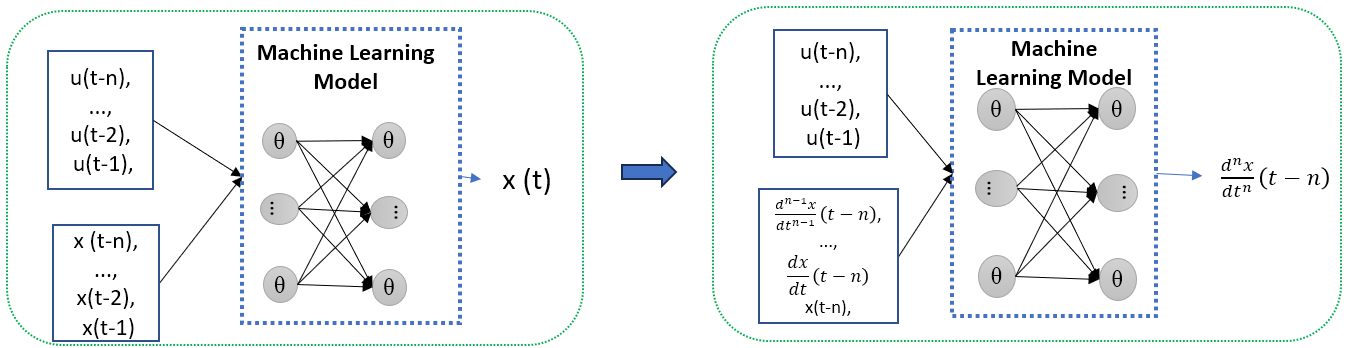}
	\caption{Using machine learning model to learn the short-term dependency.}
	\label{fig:machine learning structure}
\end{figure}

To capture this short-term dependency, represented by the $\gamma$ function, within the dynamic system, we resort to a machine learning model, potentially a neural network. Specifically, its inputs comprise past system states spanning $x(t-n),\ldots,x(t-1)$ and earlier control inputs $u(t-n),\ldots,u(t-1)$, while the output is predicting the current system state $x(t)$. The detailed design of this machine learning model can be observed in the subsequent Figure \ref{fig:machine learning structure}. 

If we employ previous states $x(t-n),\ldots,x(t-1)$ to forecast the current state $x(t)$ as the left part in above Figure \ref{fig:machine learning structure}, this is feasible given that the sampling period $\delta$ is sufficiently small. Due to the diminutive nature of the sampling period, there's a pronounced resemblance between the current and the preceding state, denoted as $x(t) \approx x(t+1)$. In such circumstances, the model might bypass the dynamic behavior and merely approximate the current state to the preceding one. To circumvent this state correlation, we opt to utilize the system state $x(t-n)$ in conjunction with lower order differential values $\frac{dx}{dt}(t-n), \dots, \frac{d^{n-1}x}{dt^{n-1}}(t-n)$ as inputs, with the intent of predicting the highest order differential value $\frac{d^n x}{dt^n}(t)$.

\subsection{Model-free physical-informed deep operator network} 
\label{subsec: Our Onet}

Once the short-term dependence is captured through a machine learning model, we combine it with DeepOnet. The DeepOnet can directly formulate the mapping between the initial state, control inputs sequence, and the time steps to the current state. While DeepOnet typically requires extensive data for training, we face a limitation of data in our situation. To address this, we leverage the short-term dependence to enhance the performance of DeepOnet.

Our architecture adopts the structure of the DeepOnet framework, consisting of both a branch and a trunk network. The branch network processes the initial system state $x(0)$ together with a sequence of control inputs, initiating from $u(0)$ and extending as $u(1), u(2), \ldots, u(t),\ldots$. Concurrently, the trunk network receives the time step $t$ as input. The ultimate output of DeepOnet is the anticipated system state $\hat{x}(t)$.

The combined loss $L$, used to train our architecture, is the sum of the conventional data loss $L_{data}$ and the physics-informed loss $L_P$. The data loss $L_{data}$ arises from the difference between the actual system state $x(t)$ and the DeepOnet's estimated state $\hat{x}(t)$.

For the physics-informed loss, we employ Auto Differential \cite{neidinger2010introduction} to compute differential values $\frac{d\hat{x}}{dt}(t)$, $\frac{d^2\hat{x}}{dt^2}(t)$, $\dots$, $\frac{d^{n-1}\hat{x}}{dt^{n-1}}(t)$,$\frac{d^{n}\hat{x}}{dt^{n}}(t)$ of the projected system state $\hat{x}(t)$. The model, which learns short-term dependence, takes in the predicted state $\hat{x}(t)$, lower order differential values $\frac{d\hat{x}}{dt}(t), \frac{d^2\hat{x}}{dt^2}(t), \dots, \frac{d^{n-1}\hat{x}}{dt^{n-1}}(t)$, and control inputs $u(t+n-1),\dots, u(t)$. Subsequently, it forecasts the highest differential value $\frac{d^{n}\bar{x}}{dt^{n}}(t)$. The physics-informed loss $L_P$ is derived from the discrepancy between $\frac{d^{n}\hat{x}}{dt^{n}}(t)$ and $\frac{d^{n}\bar{x}}{dt^{n}}(t)$.

\begin{figure}[H]
	\includegraphics[scale=0.5]{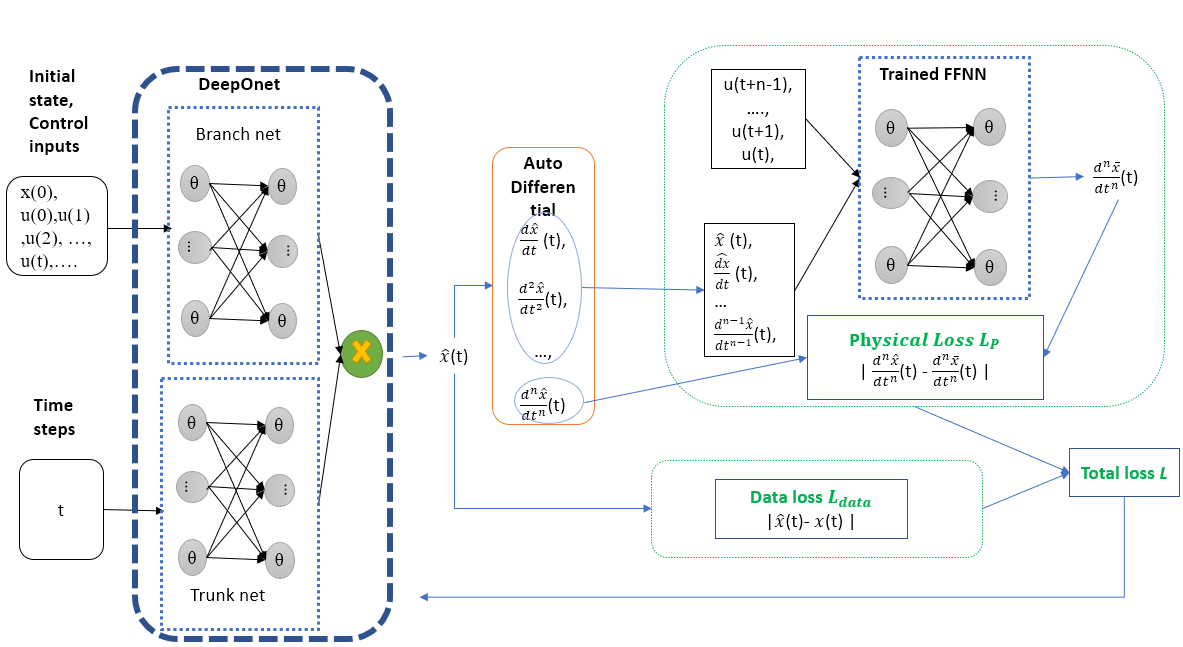}
	\caption{The proposed model-free physical-informed DeepOnet  architecture}
	\label{fig:proposed structure}
\end{figure}

Our proposed method combines the advantages of both Physics Informed Neural Networks (PINN) and Deep Operator Networks (DeepOnet) to address the limitations associated with limited system model information and a small amount of experimental data. PINN is known for its ability to work with limited data and uses a neural network to solve the differential equation-based model. However, it assumes a known model, which may not always be available in practical applications.

On the other hand, DeepOnet is a model-free technique that does not require a known model, making it suitable for cases where system model information is limited. However, it typically requires a large number of experiments to learn the system behavior accurately. By combining the strengths of both methods, our proposed technique aims to overcome the limitations of limited system model information and small amounts of experimental data, while still being able to accurately predict system behavior under different inputs or conditions.

In addition, our proposed method also incorporates our previous research on short-term dependency, which considers the influence of past states or inputs on the system's current state. This allows us to capture the memory and history of the system, further enhancing the accuracy and robustness of our approach.
\section{Result}
\label{sec:result}

In this section, we apply our method to three different systems based on ODEs to comprehensively evaluate its performance. The first, the well-known Pendulum system, is stabilized by setting the control input within a specific range. The second, the Driven Oscillator system, is made marginally unstable with a specific control input range. Both of these systems are modeled by 2nd-order ODEs. Additionally, we have tested our method on a more complex 3rd-order ODE system, the Chaotic Jerk system.

\subsection{Pendulum  System}
\label{sec:Pendulum}
The pendulum system from \cite{li2022learning} was previously discussed. It involves swinging a pendulum from a downward to an upright position. The system is defined by:

\begin{equation}
	\ddot{x}(\frac{1}{4}ml^2+I) + \frac{1}{2}ml\sin(x) = u(t)-b\dot{x}
	\label{eq: pendulum}
\end{equation}

Here, $x$ represents the pendulum angle, $\dot{x}$ its angular velocity, and $u(t)$ the control input. For this study, parameters are: $m = 1$ kg, $l = 1$ m, $I = \frac{1}{12}ml^2$, and $b = 0.01$ sNm/rad.

Below are detailed steps on how we implemented our method for the Pendulum system:

\begin{enumerate}

\item \textbf{Data Generation} \label{step:data_gen}

Data are obtained from numerical simulations generated using the Euler method.  
\begin{algorithm}
\caption{Euler method numerical simulation}
\begin{algorithmic}[1]
\State Define initial state $x(0)$, $\dot{x}(0)$
\State Input control sequence $u(0)$, $u(1)$, $u(2)$, ..., $u(t)$,...
\State Input system parameters: $m, l, g, b$
\State Set sampling period $dt=0.001s$
\For{$t = 0$ to $T-1$}
    \State $\dot{x}(t + 1) \gets \dot{x}(t) + dt \cdot \frac{u(t) - b\dot{x}(t) - \frac{1}{2} mgl\sin(x(t))}{\frac{1}{4} ml^2 + l}$
    \State $x(t+1) \gets x(t+1) + dt \cdot \dot{x}(t)$
\EndFor
\State End in $T$ step
\end{algorithmic}
\end{algorithm}

To ensure stable behavior in the pendulum system, the control input is defined as $u(t) = -c \cdot \dot{x}(t)$, where $c$ is a constant parameter. This paper addresses challenges related to limited system data. As a result, only five sets of training datasets are produced, corresponding to $c$ values from 0.35 to 0.75 with 0.1 intervals.  Furthermore, ten test datasets have been created with $c$ values ranging from 0.3 to 0.84 in increments of 0.06.

\item \textbf{Using Machine Learning Models to Learn Short-Term Dependency} \label{step:ml_model}

We are investigating the short-term dependency in the pendulum system. Since we only know that the order of the partial differential equation (\ref{eq: pendulum}) is $2$, based on Theorem \ref{theorem:n step dependency}, we can infer that the system state $x_{t}$ only depends on the most recent 2-step states and control inputs, namely $x(t-1)$, $x(t-2)$, $u(t-1)$, and $u(t-2)$.

We utilized a Feed-forward Neural Network (FFNN) to discern the short-term dependency, employing five groups of data generated as outlined in section \ref{sec:machine}. The hyper-parameters of  FFNN is shown in Table \ref{table:Network parameters}.  Upon completing the training process, our FFNN model demonstrated stability and was adept at accurately predicting the second differential value $\ddot{x}(t-1)$ of the pendulum's angle.

\item \textbf{Combining the Machine Learning Model with DeepOnet} \label{step:combine_model}

After capturing the short-term dependence on the FFNN, we integrated the FFNN with DeepOnet to devise a Physical-informed DeepOnet structure. Our architecture, tailored for the pendulum system, encompasses both a branch and a trunk network. As shown in Figure \ref{fig:proposed structure}. The branch network processes the initial system state $x(0)$ alongside the control input sequence $u(0),u(1),u(2),...,u(t),...$. Conversely, the trunk network is ftime instant: $t$. The culmination of the DeepOnet's computations is the forecasted system state, specifically $\hat{x}(t)$.  The hyper-parameters of DeepOnet are shown in Table \ref{table:Network parameters}. 
\end{enumerate}

\begin{table}[htbp]
	\centering
	\caption{Networks Hyperparameters for Three Systems}
	\includegraphics[scale=0.55]{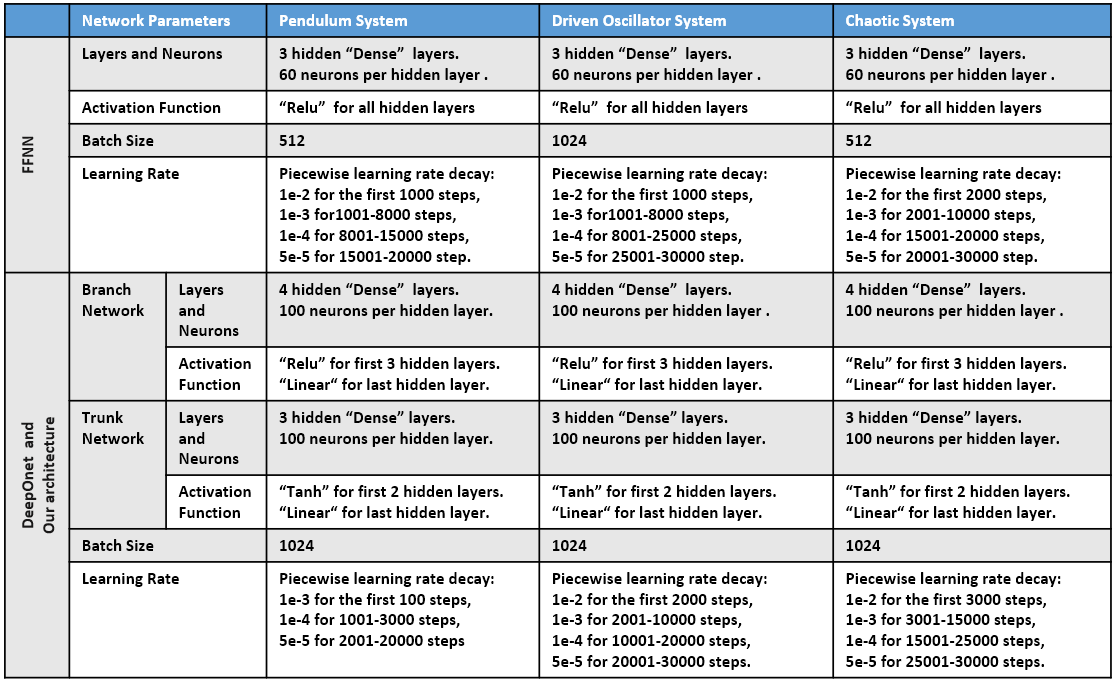}
	\label{table:Network parameters}
\end{table}

\begin{figure}[H]
	\centering
	\begin{minipage}[b]{0.48\textwidth}
		\includegraphics[width=\textwidth]{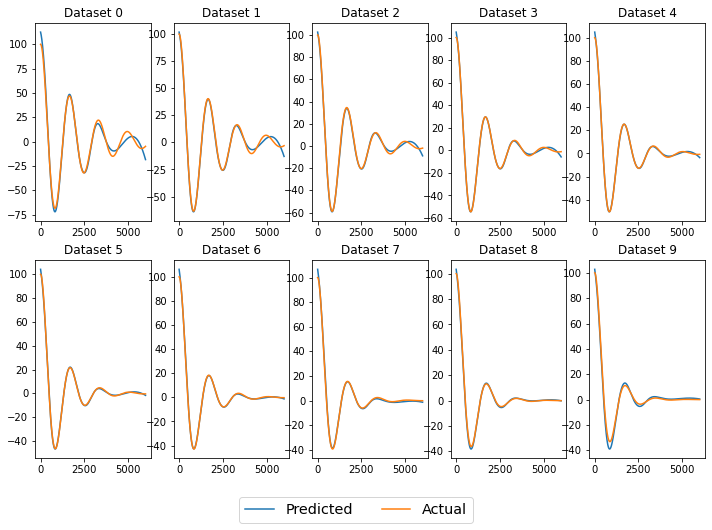}
		\caption{Test Results of DeepOnet on the Pendulum System without Noise}
		\label{fig:Pendulum_Onet_clean}
	\end{minipage}
	\hfill  % Add some horizontal spacing between the two images
	\begin{minipage}[b]{0.48\textwidth}
		\includegraphics[width=\textwidth]{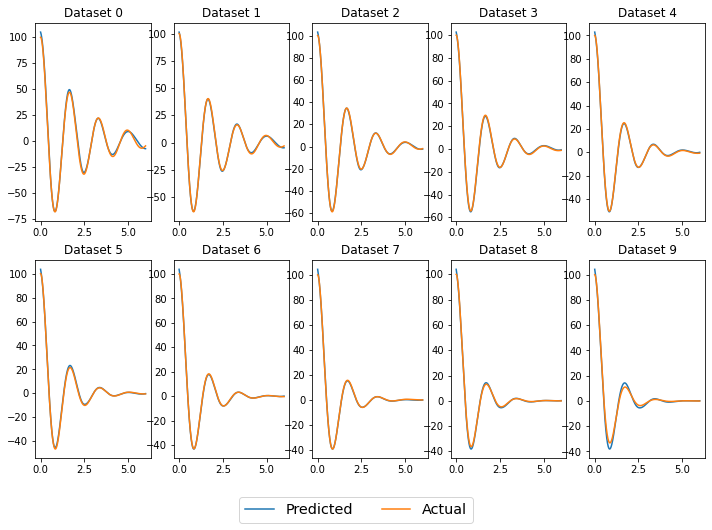}
		\caption{Test Results of Our Approach on the Pendulum System without Noise}
		\label{fig:Pendulum_our_clean}
	\end{minipage}
\end{figure}

Once our approach's network is trained, we employ 10 test datasets, generated as per Step \ref{step:data_gen}, to assess its performance. It's crucial to highlight that while these test datasets differ from the training data, their underlying control inputs are analogous to the training set. We employ DeepOnet as a benchmark for gauging the efficacy of our method. Figure \ref{fig:Pendulum_Onet_clean} and \ref{fig:Pendulum_our_clean}  provide a comparative study of the estimation results between our method and the benchmark, DeepOnet.

By comparing Figure \ref{fig:Pendulum_Onet_clean} and \ref{fig:Pendulum_our_clean}, our method showcases superior accuracy in its predictions, closely aligning with the true values throughout the displayed scenarios. Conversely, although DeepOnet captures the general trends, it presents noticeable discrepancies in specific plots, notably the first 3 datasets. This contrast highlights the refined reliability and precision of our technique when compared to the conventional DeepOnet. A more quantified differentiation is observed when examining the average mean absolute error (MAE). With an MAE of 17.98 for DeepOnet in contrast to a significantly lower 6.23 for our method, it's evident that our approach offers enhanced performance.

\textbf{Implementation with Noisy Data}

The results presented thus far pertain to noise-free data. For a more comprehensive evaluation, we tested our approach using data with measurement noise, which is often present in real-world systems. This noisy data is generated using the Euler method, as described in step \ref{step:data_gen}, with the introduction of Gaussian noise to simulate measurement inaccuracies.

\begin{figure}[htbp]
	\centering
	\includegraphics[scale=0.4]{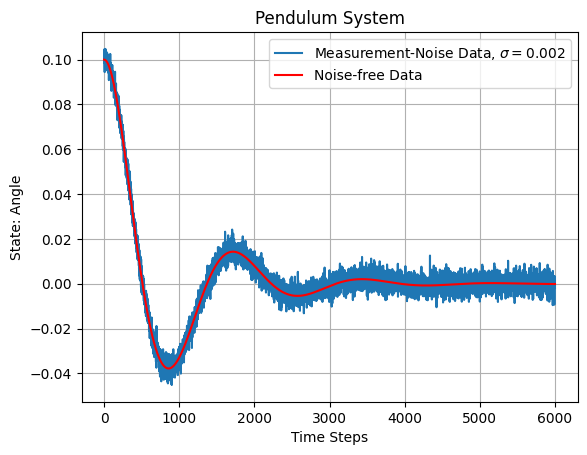}
	\caption{Comparison between noise-free data and measurement-noise data ($\mu = 0.002$) in Pendulum system}
	\label{fig: Pendulum noise vs clean}
\end{figure}

The Gaussian noise is represented by the probability density function (PDF) of a Gaussian distribution:

\begin{equation}
N(x|\mu, \sigma^2) = \frac{1}{\sqrt{2\pi\sigma^2}} e^{-\frac{(x-\mu)^2}{2\sigma^2}}
\label{eq:gaussian}
\end{equation}

where $ \mu $ denotes the mean and $ \sigma^2 $ signifies the variance. To simulate varying noise levels, we kept $ \mu = 0 $ and adjusted the value of $ \sigma^2 $. In this paper, we present the result of noise data when the variance $\sigma= 0.002$.  Figure \ref{fig: Pendulum noise vs clean} is the comparison between the noisy-free data and noisy data.  

For the noisy dataset, we first use the Savitzky-Golay filter \cite{savitzky1964smoothing} to smooth the noise data. This filter is renowned in signal processing for its robust noise elimination capabilities and for enhancing the signal's smoothness. A notable merit of the Savitzky-Golay filter is its proficiency in calculating the smoothed derivatives of the data.

\begin{figure}[H]
	\centering
	\begin{minipage}[b]{0.48\textwidth}
		\includegraphics[width=\textwidth]{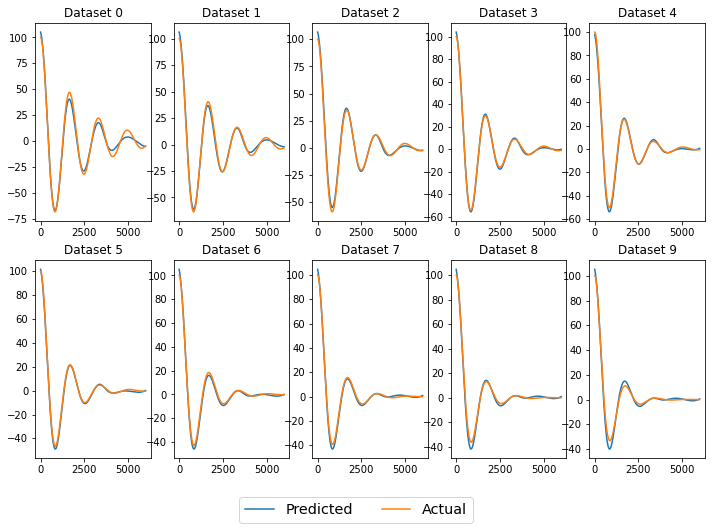}
		\caption{Test Results of DeepOnet on the Pendulum System with Measurement Noise}
		\label{fig:Pendulum_Onet_noise}
	\end{minipage}
	\hfill  % Add some horizontal spacing between the two images
	\begin{minipage}[b]{0.48\textwidth}
		\includegraphics[width=\textwidth]{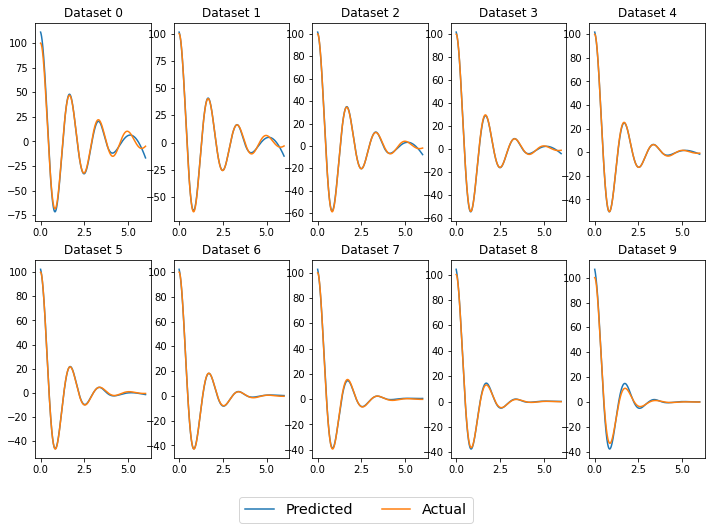}
		\caption{Test Results of Our Approach on the Pendulum System with Measurement Noise}
		\label{fig:Pendulum_our_noise}
	\end{minipage}
\end{figure}

Utilizing the procedures detailed in steps \ref{step:ml_model} and \ref{step:combine_model}, we trained the corresponding FFNN and integrated model,  which has the same parameters as the noisy-free case. To provide a tangible comparison of the efficacy of our approach against existing methods, we present Figures \ref{fig:Pendulum_Onet_noise} and \ref{fig:Pendulum_our_noise}. These visuals distinctly showcase the contrast between the results yielded by our approach and those of DeepOnet.  To quantify this performance disparity further, we calculated the MAE. Our approach yielded an MAE of 7.57, whereas DeepOnet's was 19.27.

\subsection{Driven Oscillator System}
\label{sec:Driven}
As introduced by the work in \cite{DrivenOscillator}, the Driven Oscillator System is a classic representation of an oscillator subjected to an external force. It's a foundational tool in the study of numerous physical processes, and its behavior intricately changes with variations in the driving force and inherent oscillator properties. The governing dynamics of this system can be expressed as:

\begin{equation}
	m \ddot{x} + c \dot{x} + kx = F_0u
\end{equation}

Here, $m$ characterizes the oscillator's mass, $x$ its displacement, $\dot{x}$ its velocity, and $\ddot{x}$ its acceleration. The parameters $c$ and $k$ are the damping and stiffness coefficients, respectively. The driving force's amplitude and angular frequency are represented by $F_0$ and $w$. In our specific system setup, we selected values of $m = 0.5$, $k = 50$, $c = 1$, and $F_0 = 4$.

Regarding the data for this system, it's derived using the Euler method, as referenced in Step \ref{step:data_gen}. We've generated 5 training datasets. To instill a marginally unstable behavior in the oscillator, the angular frequency $w$ was controlled between values 9 to 11 with a step of 0.5 for these datasets. Additionally, 10 test datasets were formulated with angular frequencies $w$ ranging from 8.5 to 11.2, incremented by 0.3.

\begin{figure}[H]
	\centering
	\begin{minipage}[b]{0.48\textwidth}
		\includegraphics[width=\textwidth]{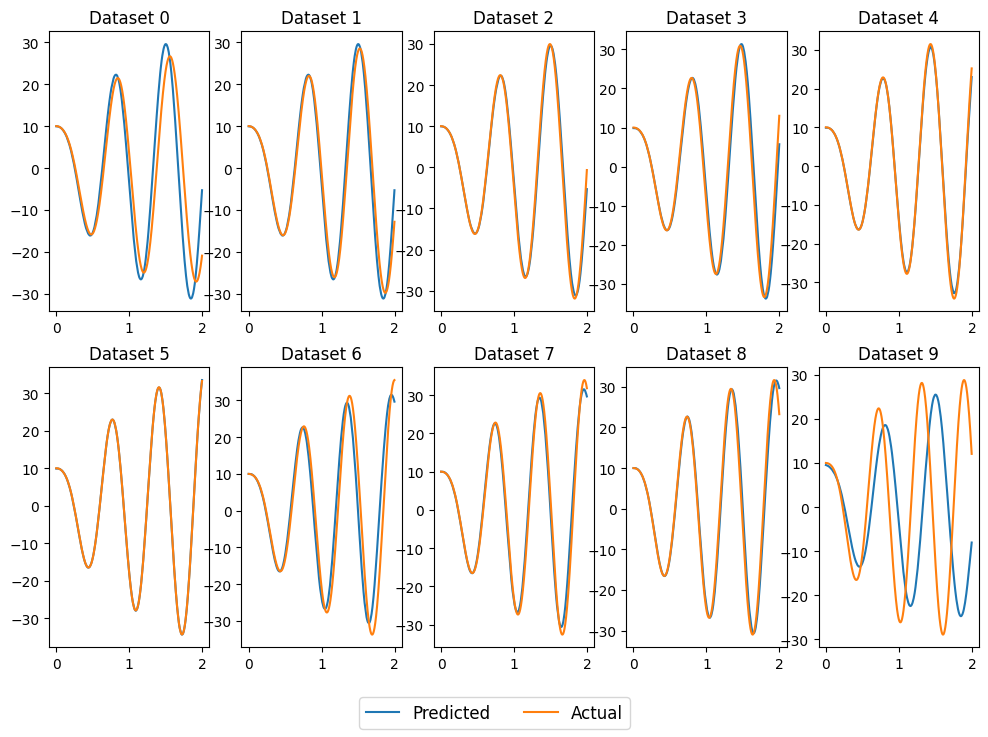}
		\caption{Test Results of DeepOnet on the Oscillator System without Noise}
		\label{fig:Oscillator_Onet_clean}
	\end{minipage}
	\hfill  % Add some horizontal spacing between the two images
	\begin{minipage}[b]{0.48\textwidth}
		\includegraphics[width=\textwidth]{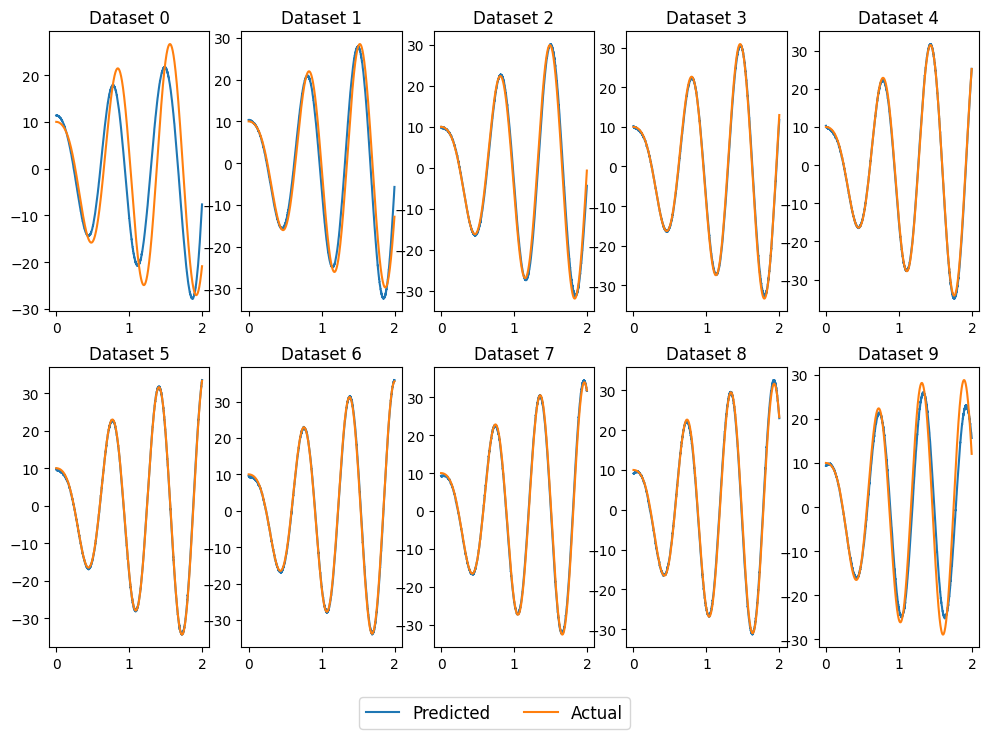}
		\caption{Test Results of Our Approach on the Oscillator System without Noise}
		\label{fig:Oscillator_our_clean}
	\end{minipage}
\end{figure}

Once the data was generated, both the FFNN and our proposed structure were trained. The hyperparameters for both the FFNN and DeepOnet, as used in our approach, can be found in Table \ref{table:Network parameters}. We evaluated the models using ten test datasets to gauge their performance. A comparative analysis of the estimation results between our method and the benchmark, DeepOnet, is presented in Figures \ref{fig:Oscillator_Onet_clean} and \ref{fig:Oscillator_our_clean}. Notably, our method demonstrated superior performance, especially for Dataset 9. To provide a numerical assessment of this performance difference, we computed the Mean Absolute Error (MAE). Our method achieved an MAE of 16.69, in contrast to DeepOnet's MAE of 36.22.

\textbf{Implementation with Noisy Data}

\begin{figure}[htbp]
	\centering
	\includegraphics[scale=0.5]{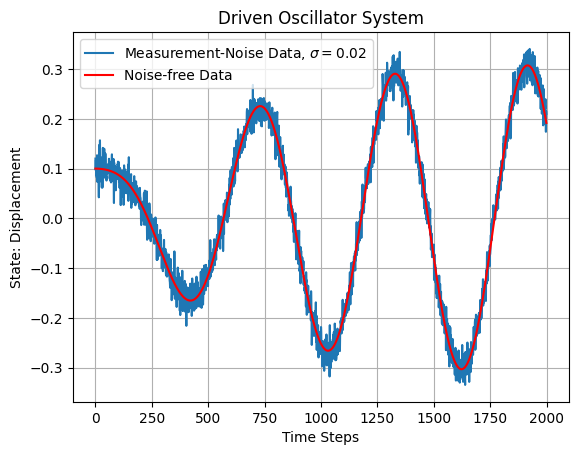}
	\caption{Comparison between noise-free data and measurement-noise data ($\mu = 0.02$) in Driven Oscillator system}
	\label{fig: Oscillator noise vs clean}
\end{figure}

We further evaluated our approach by introducing measurement noise to the data. Specifically, Gaussian noise, as defined in equation \ref{eq:gaussian}, was incorporated. The results with a noise variance, $\sigma$, of 0.01 are discussed below.  A comparative visual representation between the noise-free and noisy data is provided in Figure \ref{fig: Oscillator noise vs clean}.

\begin{figure}[H]
	\centering
	\begin{minipage}[b]{0.48\textwidth}
		\includegraphics[width=\textwidth]{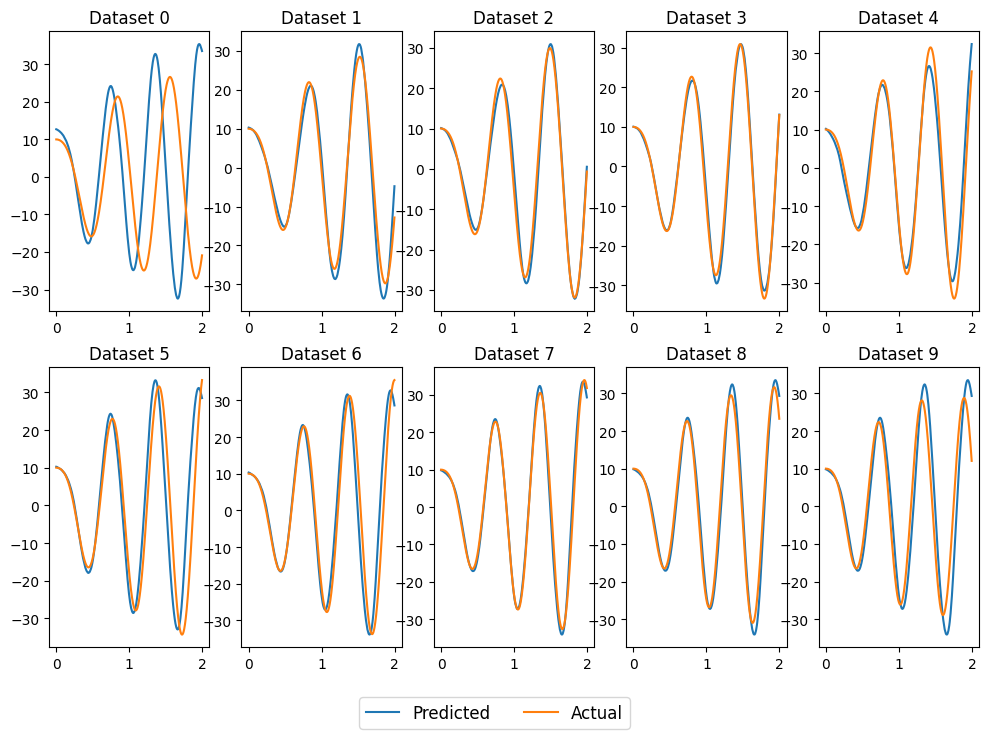}
		\caption{Test Results of DeepOnet on the Oscillator System with Measurement Noise}
		\label{fig:Oscillator_Onet_noise}
	\end{minipage}
	\hfill  % Add some horizontal spacing between the two images
	\begin{minipage}[b]{0.48\textwidth}
		\includegraphics[width=\textwidth]{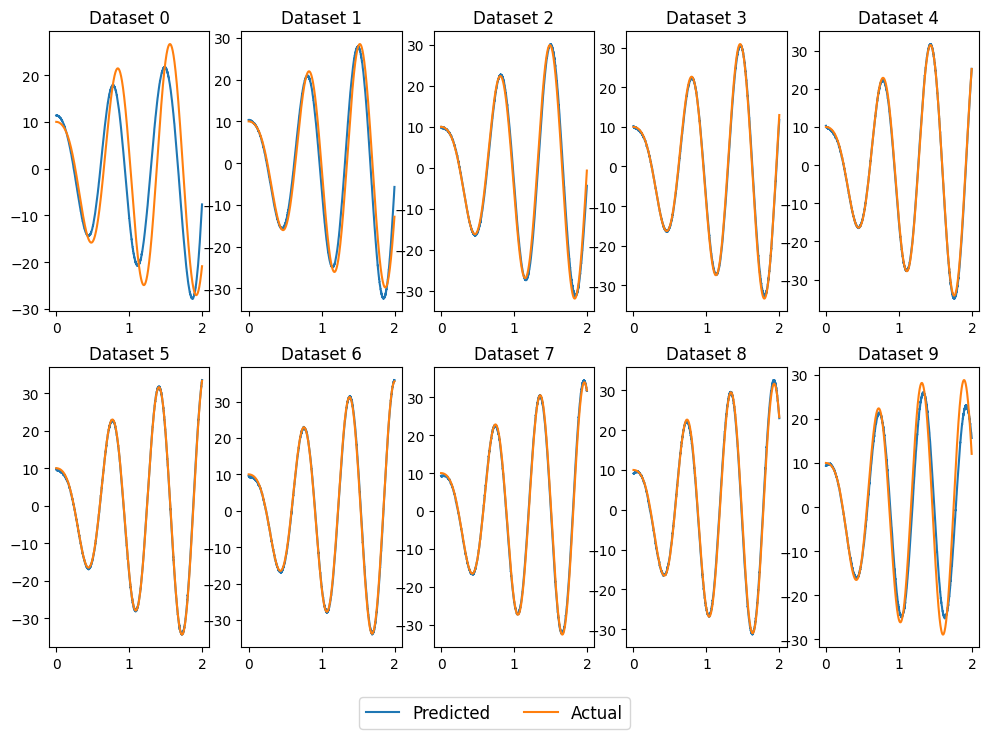}
		\caption{Test Results of Our Approach on the Oscillator System with Measurement Noise}
		\label{fig:Oscillator_our_noise}
	\end{minipage}
\end{figure}

In the context of measurement noise, the performance results are depicted in Figures \ref{fig:Oscillator_Onet_noise} and \ref{fig:Oscillator_our_noise}. While DeepOnet's predictions sometimes diverge from the actual outcomes, notably in Datasets 2, 6, 7, and 9, Our Approach more consistently mirrors the true data across the majority of the datasets. From a quantitative perspective, we determined the MAE for both methods. DeepOnet registered an MAE of 39.46, in contrast to the notably superior MAE of 19.10 achieved by our method. Even though both methods bring forth meaningful contributions, Our Approach evidently emerges as the more reliable and precise alternative in these scenarios.

\subsection{Chaotic Jerk System}
\label{sec:Jerk}
To validate the applicability of our method in handling high-order differential systems, we have chosen to implement it in a chaotic jerk system. A jerk is defined as the rate of change of acceleration, which corresponds to the third derivative of position with respect to time. In dynamical systems, a chaotic jerk system typically exhibits unpredictable and highly sensitive behavior to initial conditions. Chaotic systems, including chaotic jerk systems, are characterized by complex and non-repeating patterns of motion, known as chaotic attractors.

For our analysis, we utilize a simple chaotic jerk function described by Sprott \cite{sprott1997some}. The formula for this function is presented below:

\begin{equation}
	\dddot{x} + A\ddot{x} - \dot{x}^2 + x = u
\end{equation}

In this equation, $x$ represents the state variable, $\dot{x}$ represents its first derivative, $\ddot{x}$ represents its second derivative, and $\dddot{x}$ represents its third derivative. The parameter $A$ determines the strength of the damping term, and $u$ represents the control input.  The $u$ is further represented as $u(t)=F(t)\cdot sin(wt)$, where $F$ is the force, $w$ is frequency.

Regarding the data for this system, it's derived using the Euler method, as referenced in Step \ref{step:data_gen}. We've generated 10 training datasets. In order to have some curve behavior in the Jerk system, the $F$ is set from 1.09 to 1.9 with an interval of 0.09, and the frequency $w$ was controlled between values 2 to 2.81 with a step of 0.0.09 for these datasets. Additionally, 5 test datasets were formulated with force $F$ ranging from 1.2 to 2, incremented by 0.2, and frequencies $w$ ranging from 2 to 2.5, incremented by 0.1.

\begin{figure}[H]
	\centering
	\begin{minipage}[b]{0.48\textwidth}
		\includegraphics[width=\textwidth]{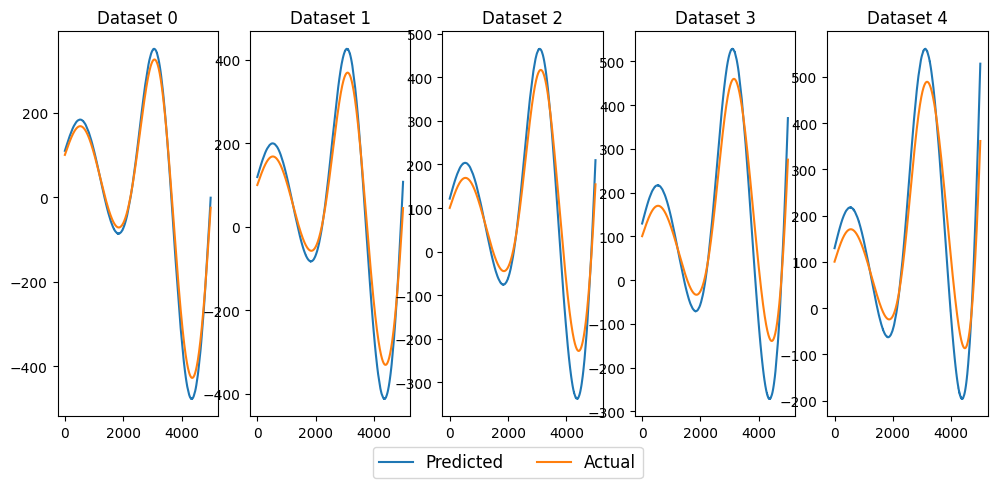}
		\caption{Test Results of DeepOnet on the Jerk System without Noise}
		\label{fig:Jerk_Onet_clean}
	\end{minipage}
	\hfill  % Add some horizontal spacing between the two images
	\begin{minipage}[b]{0.48\textwidth}
		\includegraphics[width=\textwidth]{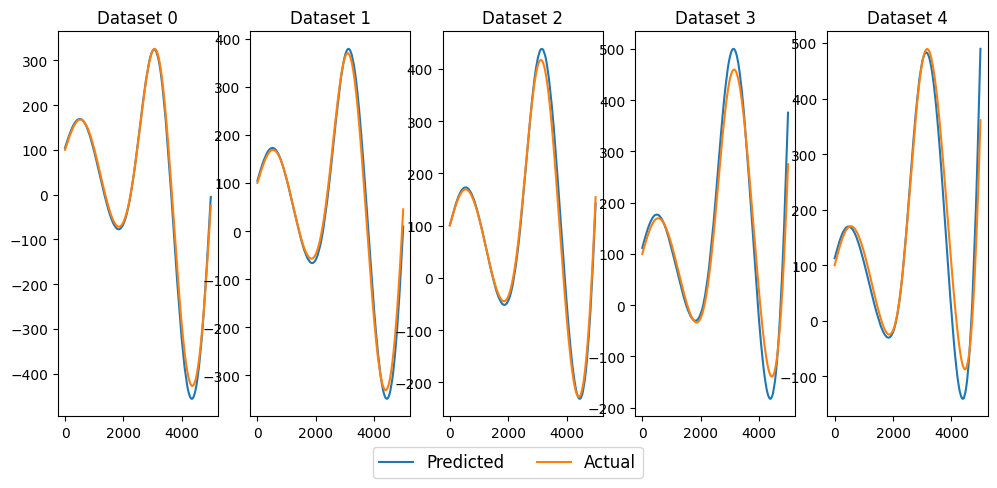}
		\caption{Test Results of Our Approach on the Jerk System without Noise}
		\label{fig:Jerk_our_clean}
	\end{minipage}
\end{figure}

Once the data was generated, both the FFNN and our proposed structure were trained. The hyperparameters for both the FFNN and DeepOnet, as used in our approach, can be found in Table \ref{table:Network parameters}. We evaluated the models using ten test datasets to gauge their performance. The estimative outcomes between our strategy and the benchmark, DeepOnet, are delineated in Figures \ref{fig:Jerk_Onet_clean} and \ref{fig:Jerk_our_clean}. Remarkably, our technique showcased enhanced accuracy across all test datasets. The MAE was also calculated in a quantified performance comparison. While our method registered an MAE of 19.8, DeepOnet recorded an MAE of 10.5. In a direct juxtaposition of results for the Jerk System without noise, our method unambiguously manifests greater precision and dependability.

\textbf{Implementation with Noisy Data}

\begin{figure}[htbp]
	\centering
	\includegraphics[scale=0.5]{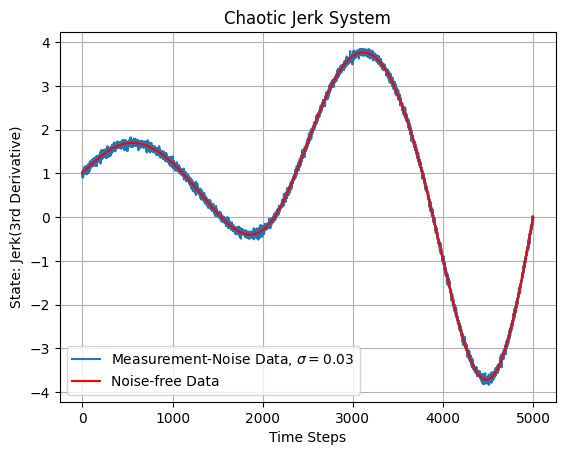}
	\caption{Comparison between noise-free data and measurement-noise data ($\mu = 0.002$) in Chaotic Jerk system}
	\label{fig: Jerk noise vs clean}
\end{figure}

We further evaluated our approach by introducing measurement noise to the data. Specifically, Gaussian noise, as defined in equation \ref{eq:gaussian}, was incorporated. Results for the noisy data, with a variance, $\sigma$, set to 0.03, are detailed. Figure \ref{fig: Jerk noise vs clean} provides a comparative visual representation between the noise-free and noisy data.

\begin{figure}[H]
	\centering
	\begin{minipage}[b]{0.48\textwidth}
		\includegraphics[width=\textwidth]{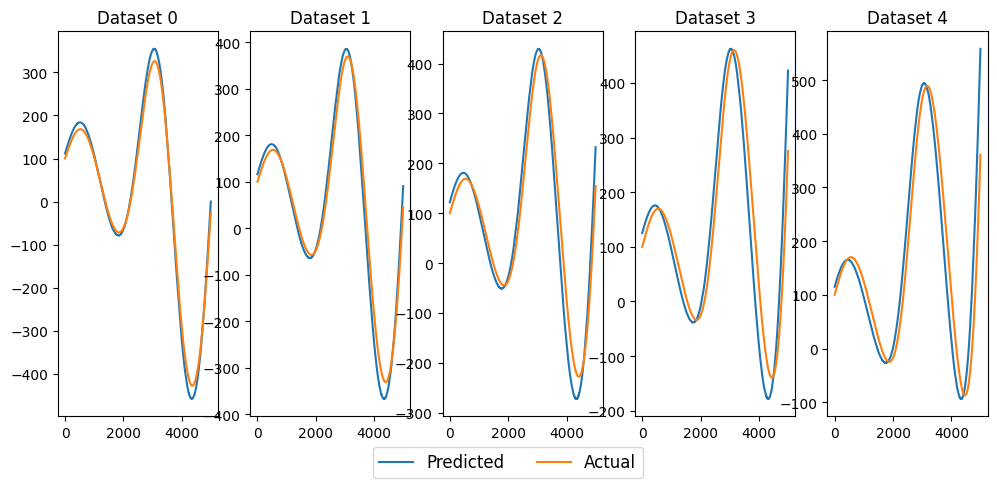}
		\caption{Test Results of DeepOnet on the Jerk System with Measurement Noise}
		\label{fig:Jerk_Onet_noise}
	\end{minipage}
	\hfill  % Add some horizontal spacing between the two images
	\begin{minipage}[b]{0.48\textwidth}
		\includegraphics[width=\textwidth]{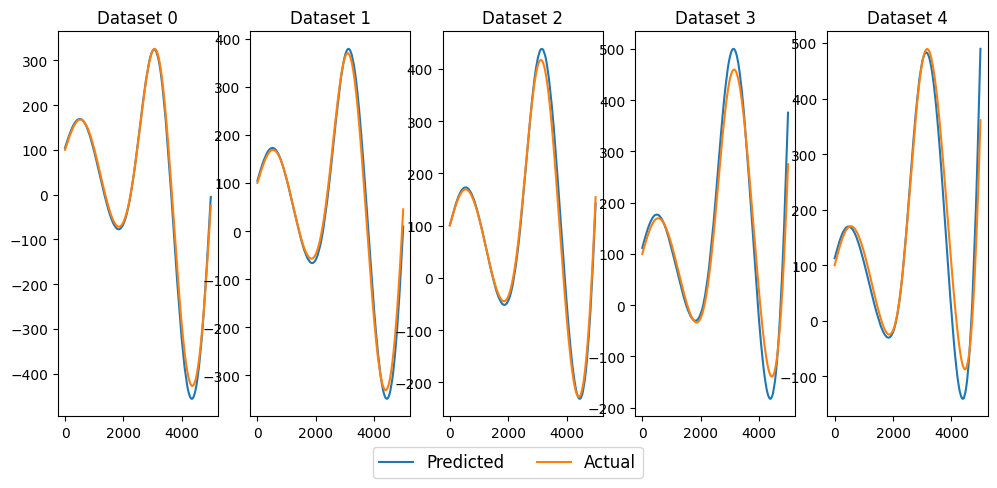}
		\caption{Test Results of Our Approach on the Jerk System with Measurement Noise}
		\label{fig:Jerk_our_noise}
	\end{minipage}
\end{figure}

Performance evaluations under the influence of measurement noise are showcased in Figures \ref{fig:Jerk_Onet_noise} and \ref{fig:Jerk_our_noise}. Across all test datasets, our approach consistently outperforms DeepOnet. To provide a numerical perspective on this performance difference, we calculated the Mean Absolute Error (MAE). While DeepOnet yielded an MAE of 22.2, our method achieved a lower value of 18.7.

\section{Conclusion}

In the rapidly evolving landscape of dynamic systems analysis, the challenges of sparse data and incomplete system modeling continue to pose significant hurdles. In our investigation, we successfully tackled these challenges with a novel approach grounded in a data-driven model-free physical-informed Deep Operator Network (DeepOnet) framework. This approach distinguished itself by tapping into the short-term dependencies inherent in the limited data, thereby harnessing the power of a surrogate machine-learning model. This surrogate model then seamlessly integrated with DeepOnet, infusing it with critical physical insights.

Our evaluation was methodical, encompassing three unique dynamic systems, each characterized by its own set of complexities, behaviors, and underlying ODE foundations. Across this spectrum of systems, our method consistently demonstrated superiority over the original DeepOnet, not only for pristine, noise-free data but also when the data was tainted with measurement noise. This resilience in the face of noise underscores the robustness of our methodology, highlighting its potential applicability in real-world scenarios fraught with uncertainties.

In conclusion, our endeavors in this study mark a significant stride in the realm of offline state estimation for dynamic systems. By effectively navigating the twin challenges of limited data and model ambiguity, we've showcased the potential of our approach. As the quest for more efficient, accurate, and versatile state estimation techniques persists, our research stands as a beacon, illuminating the path forward and underscoring the promise that data-driven methodologies hold in bridging the knowledge gaps of today.

\label{sec:conclu}
\bibliographystyle{elsarticle-num}
\bibliography{references}

\end{document}